%% file: main.tex
\documentclass[runningheads]{llncs}
\usepackage[english]{babel}
\usepackage[T1]{fontenc}
\usepackage{microtype}
\usepackage{graphicx}
\usepackage{amsmath}   %
\usepackage{amssymb}   %
\usepackage{wasysym}
\usepackage{orcidlink}
\usepackage{booktabs}
\usepackage{csquotes}
\usepackage{comment}
\usepackage{paralist}

\usepackage{xcolor}
\definecolor{codegray}{rgb}{0.5,0.5,0.5}
\definecolor{codeblue}{rgb}{0.1,0.2,0.6}
\definecolor{codegreen}{rgb}{0.1,0.5,0.1}

\usepackage{listings}
\lstdefinelanguage{json}{
    morestring=[b]",
    morestring=[d]',
    literate=
      *{:}{{{\color{codeblue}{:}}}}{1}
       {,}{{{\color{codeblue}{,}}}}{1}
       {\{}{{{\color{codeblue}{\{}}}}{1}
       {\}}{{{\color{codeblue}{\}}}}}{1}
       {[}{{{\color{codeblue}{[}}}}{1}
       {]}{{{\color{codeblue}{]}}}}{1},
}

\usepackage{todonotes}

\usepackage{tikz}
\usetikzlibrary{positioning,arrows.meta,calc,fit,backgrounds,shapes.geometric}
\usepackage[capitalize]{cleveref}

\usepackage{hyperref}

\title{Lightweight Zero Trust via Automotive SDN}

\author{%
        Friedrich Wiemer\inst{1} \orcidlink{0000-0003-2998-6777}, and
        Florian Wagner\inst{2} %
}
\authorrunning{F. Wiemer and F. Wagner}

\institute{%
        Robert Bosch GmbH, Stuttgart, Germany\\
        friedrich.wiemer@de.bosch.com \and
        ETAS GmbH, Stuttgart, Germany\\
        florian.wagner3@etas.com
}

\begin{document}
\maketitle              %
\begin{abstract}
Zonal in-vehicle networks ship Ethernet, MACsec, and TSN, but treat the network itself as trusted: once configured at the factory, there is no standardized runtime way to easily revoke access, rotate keys, or contain a compromised ECU.
Zero Trust Architecture targets exactly that gap, yet existing automotive ZTA proposals bolt on dedicated infrastructure that duplicates the SDN management plane already required to enable SDVs.
Thus, ZTA is not yet adopted in the automotive domain, and the question remains: can we do better?
We answer this in two steps.

\emph{Step~1} analyses what Open Alliance TC17~v1.0 MACsec/MKA with pre-shared CAKs already provides in terms of NIST SP~800-207 ZTA tenets.
\emph{Step~2} adds CORECONF/YANG management as proposed in Open Alliance TC19, maps the SDN Controller and Agents one-to-one onto NIST's PE, PA, and PEP.
We then instantiate this with two YANG-based mechanisms: a network-access-control flow and a key-management scheme.

The result fully covers five and two partially of the seven tenets with no ZTA-specific infrastructure added.

\keywords{Zero Trust Architecture \and Software-Defined Networking \and CORECONF \and YANG \and In-Vehicle Network Security}
\end{abstract}

\input{introduction}
\input{background}
\input{step1}
\input{step2}
\input{conclusion}

\begin{credits}
\end{credits}

\bibliographystyle{splncs04}
\bibliography{references}

\end{document}

%% file: introduction.tex
\section{Introduction}\label{sec:introduction}

The automobile is in the middle of an architectural reinvention.
Domain-oriented electrical/electronic (E/E) architectures, with their heterogeneous patchwork of CAN, LIN, FlexRay, and Automotive Ethernet networks stitched together by central gateways, are giving way to \emph{zonal} architectures in which a small number of high-performance vehicle computers communicate with zone controllers over a switched Automotive Ethernet backbone~\cite{whitepaper}.
With this shift, the in-vehicle network (IVN) inherits the full stack of enterprise-grade mechanisms that the Ethernet ecosystem provides:
VLAN segmentation, Time-Sensitive Networking (TSN) for deterministic real-time traffic, and IEEE~802.1AE MACsec for line-rate link-layer cryptographic protection~\cite{ieee8021ae}.
At the same time, regulators have closed the door on shipping vehicles without a credible cybersecurity story:
UN~R155 mandates a certified Cybersecurity Management System over the entire vehicle lifecycle~\cite{unece_r155}, and ISO/SAE~21434 codifies the engineering process expected to satisfy it~\cite{iso21434}.

Despite this technological and regulatory momentum, the IVN itself is still treated as a \emph{trusted} medium.
Once an ECU has been assembled at the factory, it is implicitly authorised to source and consume any traffic its configuration allows for the remainder of a 15-year-plus service life -- a lifetime over which the threat landscape, the software stack, and in the future even the supplier ecosystem of a vehicle will change repeatedly.
The fragility of this assumption has been demonstrated repeatedly, e.g., recently by \emph{CAN-injection} attacks against modern vehicles, which show that physical access to a peripheral wiring harness is a realistic threat and suffices to impersonate trusted ECUs to command safety-critical functions.\footnote{%
  See, e.g., Tindell's analysis of the 2022/2023 keyless-theft CAN-injection campaigns against several OEMs.
}
Recent surveys confirm that Automotive Ethernet inherits a comparable attack surface unless its security features are actively used~\cite{devincenzi2024}.
The signal-oriented countermeasure used today, Secure On-Board Communication~(SecOC), scales linearly with the number of authenticated signals and is already approaching the limit of what OEMs can configure and maintain by hand as signal counts grow exponentially with every vehicle generation.\footnote{
  As argued, e.g., in Volvo Cars' keynote at the Automotive Ethernet Congress 2024.
}
What is missing is not another security control, but a \emph{systematic, standards-based} framework that turns the IVN from a statically trusted medium into a continuously verified network.

Zero Trust Architecture (ZTA), as formalised by NIST~SP\,800-207~\cite{nist800207}, is exactly such a framework:
its guiding principle of \enquote{never trust, always verify} replaces perimeter trust with per-request authentication, authorisation, and policy enforcement.
Applying ZTA to the IVN is therefore an obvious target, and several recent proposals have explored the design space, both for in-vehicle and vehicle-to-edge scenarios~\cite{haeckel2022,anderson2023,shipman2024,iftikhar2025}.
A common pattern in these works, however, is to introduce a \emph{dedicated} Zero Trust control plane -- additional policy engines, agents, and identity infrastructure -- placed alongside the networking and SDV management stack that the OEMs are already building.
The result is duplication of mechanism, duplication of integration effort, and, in practice, yet missing adoption.
The question we ask in this paper is therefore not \emph{whether} ZTA should be brought into the vehicle, but \emph{how lightly} this can be done, if one is willing to reuse infrastructure that the zonal, software-defined vehicle requires anyway.

\paragraph{Two-step approach.}
Our central claim is that a meaningful Zero Trust posture for the IVN can be reached \emph{incrementally}, using only standards and infrastructure that are already on the automotive roadmap, in two separated steps.

\emph{Step~1 -- what is achievable today.}
Automotive Ethernet with MACsec/MKA using pre-shared CAKs, as standardised by Open~Alliance TC17~\cite{oa_tc17_macsec,oa_tc17_mka}, combined with VLAN isolation and TSN, already delivers a surprising fraction of the NIST ZTA tenets:
per-link cryptographic authentication and confidentiality, deny-by-default at the link layer, and strong segmentation -- all from static, design-time configuration, with no additional runtime infrastructure.

\emph{Step~2 -- what CORECONF/YANG-managed SDN adds tomorrow.}
Layering a CORECONF/YANG management plane (in the spirit of Open~Alliance TC19) on top of Step~1 closes the remaining ZTA gaps -- dynamic policy distribution, centralised key lifecycle management, and runtime reconfiguration in response to incidents -- and does so by mapping the SDN Controller and on-device Agents one-to-one onto NIST's Policy Engine, Policy Administrator, and Policy Enforcement Points.

In other words, the SDN control plane \emph{is} the ZTA control plane; no separate Zero Trust infrastructure is required.

This split matters in practice.
Step~1 can be deployed, resp.\ is already under deployment, on vehicles entering production within the current development cycle, using exclusively static configuration that fits the existing release process.
Step~2 unlocks standardized, dynamic capabilities -- key rotation, ECU revocation, on-the-fly policy updates -- that long-lived, over-the-air-updated vehicles demand, but only once a YANG-based SDN management plane is available.
Treating the two as a single architectural roadmap, rather than competing alternatives, lets OEMs realise immediate security benefits without precluding the more dynamic posture that the Software-Defined Vehicle will eventually require.

\paragraph{Contributions.}
Concretely, this paper makes the following contributions:
\begin{compactenum}
  \item A systematic analysis of which NIST~SP\,800-207 tenets can be satisfied at the IVN layer using only MACsec/MKA with pre-shared CAKs, VLANs, and TSN (\emph{Step~1}).
  \item A one-to-one mapping of a CORECONF/YANG-based automotive SDN architecture onto the NIST ZTA reference components, showing that the SDN management plane can serve as the Zero Trust control plane without additional infrastructure (\emph{Step~2}).
  \item A gap analysis that makes explicit which ZTA requirements are already met by Step~1 and which strictly require the dynamic capabilities introduced in Step~2.
  \item Two concrete YANG-based mechanisms that instantiate Step~2: a network-access-control flow that replaces the
    802.1X/EAP/RADIUS stack with a CORECONF-native interaction, and a key-management scheme that standardises the MACsec CAK lifecycle across the vehicle.
\end{compactenum}
Taken together, these contributions cover six of the seven NIST ZTA tenets at the network layer without introducing any ZTA-specific infrastructure beyond what an SDN-managed zonal vehicle already needs.

\paragraph{Paper structure.}
\Cref{sec:background} reviews the relevant background on ZTA, MACsec/MKA, and CORECONF/YANG-based SDN, and positions the paper against prior automotive ZTA work.
\Cref{sec:step1} develops Step~1 and analyses the ZTA properties of static MACsec deployments.
\Cref{sec:step2} introduces the SDN-managed Step~2 and details the two YANG-based mechanisms.
\Cref{sec:conclusion} concludes our paper.

%% file: background.tex
\section{Background and Related Work}\label{sec:background}

This section recalls relevant background and related work for the remaining work.

\subsection{Zero Trust Architecture}\label{sec:bg-zta}

NIST SP\,800-207~\cite{nist800207} is the reference for Zero Trust Architecture (ZTA).
It replaces perimeter-based trust with three guiding principles: \begin{inparaenum}
    \item verify explicitly,
    \item enforce least privilege, and
    \item assume breach.
\end{inparaenum}
A concrete ZTA deployment is decomposed into three logical components:
\begin{inparaenum}
    \item a \emph{Policy Engine}~(PE) that decides whether a given subject may access a given resource,
    \item a \emph{Policy Administrator}~(PA) that translates those decisions into actionable configuration, and
    \item \emph{Policy Enforcement Points}~(PEPs) that gate the data plane accordingly.
\end{inparaenum}
The seven tenets of \S2.1 of the standard %
provide the evaluation rubric we use throughout the paper.
The companion implementation guide NIST SP\,1800-35~\cite{nist1800_35} and the CISA Zero Trust Maturity Model~\cite{cisa_ztmm} refine these tenets into deployment maturity levels but do not change the underlying model.
A comprehensive survey of enterprise ZTA deployments is given by Syed et al.~\cite{syed2022}; Ramezanpour and Jagannath~\cite{ramezanpour2022} provide a methodologically close mapping of the same NIST components onto 5G/O-RAN, which inspired the SDN mapping in \Cref{sec:step2}.

\subsection{MACsec, MKA, and the Automotive TC17 Profile}\label{sec:bg-macsec}

IEEE\,802.1AE (MACsec)~\cite{ieee8021ae} provides authenticated encryption at the Ethernet link layer between adjacent peers.
The associated \emph{MACsec Key Agreement} (MKA) protocol, defined as part of IEEE\,802.1X~\cite{ieee8021x}, derives short-lived \emph{Secure Association Keys} (SAKs) from a longer-lived \emph{Connectivity Association Key} (CAK) shared by the members of a connectivity association (the group into which MACsec peers are organized).
The CAK itself can be provisioned in two ways: dynamically through an EAP exchange (e.g., EAP-TLS~\cite{rfc9190}), or statically as a pre-shared key (PSK).
Open Alliance TC17~v1.0~\cite{oa_tc17_macsec,oa_tc17_mka} assumes the latter for the \emph{Automotive MACsec Profile}:
each MACsec-protected link in the vehicle has a CAK assigned at production time, and MKA runs peer-to-peer without any authentication server.
This profile is the foundation of Step~1 (\Cref{sec:step1}).

Lauser et al.~\cite{lauser2024} systematically compare SecOC, MACsec, IPsec, and TLS for in-vehicle use; we adopt their conclusion that MACsec is the natural \emph{link-layer} security primitive for the zonal Ethernet backbone, leaving end-to-end primitives like SecOC and TLS for application-layer concerns out of scope.
De Vincenzi et al.~\cite{devincenzi2024} provide the threat-landscape reference for Automotive Ethernet and motivate the need for active use of these mechanisms.
Wiemer et al.~\cite{wiemer2025} show how the same MACsec/MKA construction extends to CAN\,XL (\enquote{CANsec}), which makes the mechanisms of Step~2 applicable beyond the Ethernet backbone.

\subsection{SDN and CORECONF/YANG for In-Vehicle Networks}\label{sec:bg-sdn}

Software-Defined Networking~(SDN), since OpenFlow~\cite{mckeown2008} and the survey by Kreutz et al.~\cite{kreutz2015}, separates a programmable control plane from packet-forwarding data planes.
In enterprise and carrier networks, the de-facto management toolchain has converged on YANG-modelled configuration~\cite{rfc7950} carried by NETCONF~\cite{rfc6241} or RESTCONF~\cite{rfc8040}, with~\cite{claise2019} as the standard reference.
Neither protocol is a good fit for in-vehicle ECUs: both assume XML/JSON parsing and TCP/TLS, which is incompatible with the footprint of a typical peripheral ECU.
CORECONF~\cite{coreconf} closes this gap by re-using YANG as the data model, but encoding instances in CBOR~\cite{rfc9254} and transporting them over CoAP/DTLS.
Bhat et al.~\cite{bhat2023} benchmark CORECONF against NETCONF and RESTCONF and show order-of-magnitude reductions in message size and ROM footprint on constrained devices.

The application of SDN inside the vehicle was opened up by H{\"a}berle et al.~\cite{haeberle2020}, who proposed SDN as the softwarisation lever for automotive E/E architectures.
H{\"a}ckel et al.~\cite{haeckel2020,haeckel2022} extended this line of work to secure SDN-managed TSN, including an analysis of trust zones inside the vehicle, and Nam et al.~\cite{nam2021} used SDN to simplify TSN stream reservation.
Open\,Alliance TC19 is currently consolidating these ideas into an automotive SDN profile; we are convinced it should leverage the existing CORECONF/YANG management plane and treat it as the substrate of Step~2.

\subsection{Threat Model}\label{sec:bg-threat}

We assume a Dolev--Yao adversary with physical access to the in-vehicle wiring harness: any link can be tapped, frames can be observed, modified, replayed, or injected, and an attacker may insert a rogue node on any segment.
On top of this network-level capability we admit one fully compromised peripheral ECU whose software stack is under attacker control, but whose HSM-resident keys remain non-extractable.

In particular, we assume the following:
\begin{inparaenum}
    \item Every ECU carries an HSM with non-extractable key storage and a key-wrap primitive, as the vast majority of current-generation automotive microcontrollers do;
    \item Every ECU holds a unique certificate and private key, injected out of band before or at production together with all other long-term key material, plus a path along which the vehicle computer obtains revocation information -- several OEMs already operate such a per-ECU PKI for diagnostics, component protection, and V2X provisioning;
    \item A CORECONF/YANG management plane exists in the vehicle for TSN, QoS, VLAN, and MACsec configuration irrespective of Zero Trust; and
    \item The vehicle computer hosting the SDN controller and the per-node HSMs are trusted -- their compromise, physical attacks on the controller, and side-channel or fault-injection attacks on cryptographic accelerators are out of scope.
\end{inparaenum}

\subsection{Related Work and Positioning}\label{sec:bg-related}

Three threads of prior work are directly adjacent.
\begin{inparaenum}
    \item \cite{shipman2024} is the closest match: they propose a ZTA for automotive networks, but realise it as a dedicated policy infrastructure alongside the existing communication stack; \cite{anderson2023,huang2024} apply ZTA to connected vehicles and platoon control respectively, both outside the Ethernet backbone, and all three treat ZTA as an additional layer rather than as a re-use of the SDN management plane already required for TSN, MACsec, and VLAN configuration.
    \item On the SDN-managed MACsec side, \cite{choi2018} replace standard MKA with a centralised, multi-hop key exchange; we take the opposite stance and keep MKA standards-conformant, leaving the SDN controller responsible only for CAK lifecycle and policy via CORECONF/YANG.
    \item Outside automotive, SDN-to-ZTA mappings have been explored for enterprise networks~\cite{guo2023}, smart cities~\cite{iftikhar2025}, and ML-augmented SDN deployments~\cite{bashaa2025,liang2024}; these confirm the methodology but do not address production key provisioning, constrained ECUs, vehicle lifetimes, or co-existence with TSN.
\end{inparaenum}

%% file: step1.tex
\section{Step~1: MACsec based Zero Trust}\label{sec:step1}

Analysing \enquote{Zero Trust in the vehicle} separately, leads one to reach for new infrastructure: identity servers, dedicated policy engines, certificate hierarchies tailored to ECUs.
In our first step, we instead look at a ZTA in the context of the existing system:
\emph{how much of NIST's Zero Trust Architecture is already achieved by the security mechanisms that today's vehicles carry on every Ethernet link?}
The answer, as this section will show, is \enquote{more than is usually assumed} -- but, equally importantly, \enquote{not enough}.
The resulting gap then motivates our next Step~2.

\subsection{System Model}\label{sec:step1-system}

The Step~1 system is a stripped zonal Ethernet backbone, for the ease of exposition.
A Vehicle Computer~(VC) connects to a small number of Zone Controllers~(ZCs), each ZC in turn aggregating the sensors and actuators that happen to sit in its physical proximity.
Every point-to-point Ethernet link -- both VC$\leftrightarrow$ZC and ZC$\leftrightarrow$peripheral -- carries exactly one MACsec connectivity association.
The associated CAK is provisioned out-of-band at production time, in line with \cite{oa_tc17_macsec,oa_tc17_mka}, and stored in the ECU's HSM.
At power-on, MKA runs peer-to-peer on each link and distributes fresh, short-lived SAK; thereafter every frame on the wire is MACsec-protected.
VLAN assignments are written into the switch's static configuration at the same time as the CAKs.

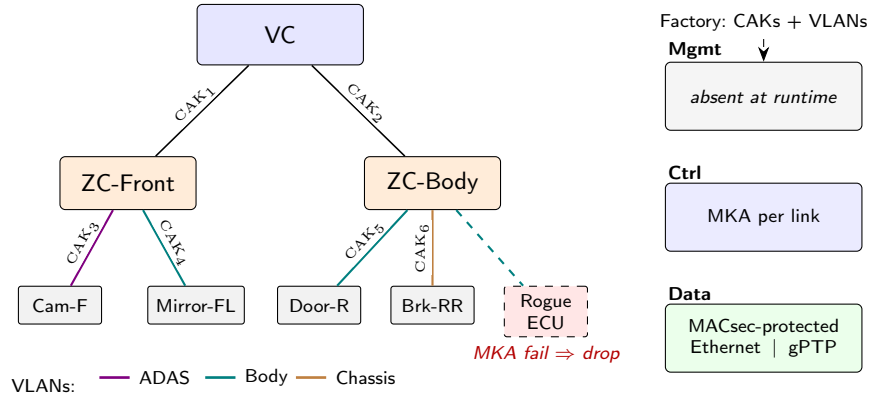
\begin{figure*}[t]
\centering
\begin{tikzpicture}[
    font=\sffamily\footnotesize,
    >={Stealth[length=2mm]},
    hpc/.style   ={draw, rounded corners=2pt, fill=blue!10, minimum width=22mm, minimum height=8mm, align=center},
    zc/.style    ={draw, rounded corners=2pt, fill=orange!15, minimum width=18mm, minimum height=7mm, align=center},
    ecu/.style   ={draw, rounded corners=1pt, fill=gray!10, minimum width=11mm, minimum height=5mm, font=\sffamily\scriptsize, align=center},
    rogue/.style ={draw, dashed, rounded corners=1pt, fill=red!10, minimum width=11mm, minimum height=5mm, font=\sffamily\scriptsize, align=center},
    macsec/.style={line width=0.6pt},
    lane/.style  ={draw, rounded corners=2pt, minimum width=26mm, minimum height=9mm, align=center, font=\sffamily\scriptsize},
    lanelbl/.style={font=\sffamily\scriptsize\bfseries, anchor=south west, inner sep=1pt},
    keytag/.style={font=\sffamily\scriptsize\itshape, fill=white, inner sep=1pt},
    vlanA/.style ={draw=violet, line width=0.8pt},
    vlanB/.style ={draw=teal,   line width=0.8pt},
    vlanC/.style ={draw=brown,  line width=0.8pt},
]
\begin{scope}[local bounding box=topo]

  \node[hpc] (hpc) at (0,0) {VC};

  \node[zc, below left=12mm and 0mm of hpc] (zcf) {ZC-Front};
  \node[zc, below right=12mm and 0mm of hpc] (zcb) {ZC-Body};

  \draw[macsec] (hpc) -- node[keytag, sloped, above] {\tiny $\mathrm{CAK}_{1}$} (zcf);
  \draw[macsec] (hpc) -- node[keytag, sloped, above] {\tiny $\mathrm{CAK}_{2}$} (zcb);

  \node[ecu, below=10mm of zcf, xshift=-9mm] (f1) {Cam-F};
  \node[ecu, below=10mm of zcf, xshift= 9mm] (f2) {Mirror-FL};
  \node[ecu, below=10mm of zcb, xshift=-15mm] (b1) {Door-R};
  \node[ecu, below=10mm of zcb, xshift=  0mm] (b2) {Brk-RR};
  \node[rogue, below=10mm of zcb, xshift= 15mm] (b3) {Rogue\\ECU};

  \draw[macsec, vlanA] (zcf) -- node[keytag, sloped, above] {\tiny $\mathrm{CAK}_{3}$} (f1);   %
  \draw[macsec, vlanB] (zcf) -- node[keytag, sloped, above] {\tiny $\mathrm{CAK}_{4}$} (f2);   %
  \draw[macsec, vlanB] (zcb) -- node[keytag, sloped, above] {\tiny $\mathrm{CAK}_{5}$} (b1);   %
  \draw[macsec, vlanC] (b2) -- node[keytag, sloped, above] {\tiny $\mathrm{CAK}_{6}$} (zcb);   %
  \draw[macsec, vlanB, dashed] (zcb) -- (b3); %

  \node[font=\sffamily\scriptsize\itshape, red!70!black, below=0mm of b3, align=center]
    {MKA fail $\Rightarrow$ drop};

  \node[font=\sffamily\scriptsize, anchor=north west] (leg) at ($(f1.south west)+(-2mm,-6mm)$) {VLANs:};
  \draw[vlanA, line width=1pt] ($(leg.east)+(2mm,1mm)$) -- ++(4mm,0)
    node[right, font=\sffamily\scriptsize] {ADAS};
  \draw[vlanB, line width=1pt] ($(leg.east)+(16mm,1mm)$) -- ++(4mm,0)
    node[right, font=\sffamily\scriptsize] {Body};
  \draw[vlanC, line width=1pt] ($(leg.east)+(28mm,1mm)$) -- ++(4mm,0)
    node[right, font=\sffamily\scriptsize] {Chassis};
\end{scope}

\begin{scope}[shift={($(topo.east)+(18mm,-2mm)$)}, local bounding box=planes]
  \node[lane, fill=gray!8]   (mgmt) at (0, 1.6) {\textit{absent at runtime}};
  \node[lane, fill=blue!8]   (ctrl) at (0, 0.0) {MKA per link};
  \node[lane, fill=green!8]  (data) at (0,-1.6) {MACsec-protected\\Ethernet\;\;|\;\;gPTP};

  \node[lanelbl] at (mgmt.north west) {Mgmt};
  \node[lanelbl] at (ctrl.north west) {Ctrl};
  \node[lanelbl] at (data.north west) {Data};

  \node[font=\sffamily\scriptsize, anchor=south, align=center] (fac) at ($(mgmt.north)+(0,3mm)$)
    {Factory: CAKs + VLANs};
  \draw[->, dashed] (fac.south) -- (mgmt.north);
\end{scope}

\end{tikzpicture}
\caption{Step~1 at a glance.
\textbf{Left:} reference zonal topology (1\,VC, 2\,ZCs, peripherals attached by physical proximity) with one MACsec connectivity association per link:
Cam-F (front camera) and Mirror-FL (front-left mirror) on ZC-Front;
Door-R (rear door module) and Brk-RR (rear-right brake actuator) on ZC-Body, alongside a rogue ECU.
Edge colour denotes the static VLAN of the attached ECU; every solid edge carries a pre-shared CAK.
ECUs of \emph{different} VLANs share the same zone controller -- zones are defined by topology, not by domain -- and the dashed red node illustrates a rogue ECU dropped by MKA because no CAK is provisioned for its link.
\textbf{Right:} the data and control planes are populated by MACsec and per-link MKA, while the management plane is \emph{absent at runtime} -- all parameters (CAKs, VLANs, ACLs) are baked in at production time.}
\label{fig:architecture}
\end{figure*}

There is no SDN controller, nor SDN agent, no RADIUS or authentication server, and no runtime management channel of any kind.
All security-relevant configuration is baked in at build time; nothing changes after the vehicle leaves the factory unless an ECU is reflashed.
\Cref{fig:architecture} captures this minimal world.

\subsection{Zero Trust Properties Achieved}\label{sec:step1-achieved}

Even this stripped-down system already satisfies, or partially satisfies, three of the seven NIST SP\,800-207 tenets (T1 to T7) at the network layer.

A MACsec receiver drops every frame that lacks a valid SecTAG and ICV.
Since MKA will only derive a SAK between peers that both prove possession of the same CAK, an ECU without the correct CAK simply cannot place legitimate traffic onto the link.
The drop is observable via the standard MACsec counters (\texttt{InPktsNoSCI}, \texttt{InPktsNotValid}, \ldots), so the attempt is auditable even though it never reaches a higher protocol layer.
In ZT vocabulary, the link itself enforces deny-by-default, without any cooperation from a higher-level policy decision point, paying towards T3, T6.

As for how T2, communication secured regardless of location, is satisfied, consider the following.
Once MKA has succeeded, all data-plane traffic on the link is authenticated and (optionally) encrypted hop by hop.
The \enquote{regardless of location} clause of NIST tenet T2 is satisfied trivially: there is no \enquote{trusted inside} on the wire -- the same MACsec frame format protects backbone and zone links alike.
A man-in-the-middle inserted on a harness segment cannot eavesdrop or modify traffic without holding the corresponding CAK, and a compromised peripheral ECU can decrypt only the connectivity associations it is itself a member of.
The blast radius of any single compromise is therefore bounded by the set of links to which that ECU holds keys -- in the worst case a single peripheral connection.

Besides, T3, asking for least-privilege per-session access, is satisfied partially by the combination of static VLANs and switch-port ACLs at the ZC.
These restrict each peripheral to the small subset of the network that its function requires.
Critically, and as \cref{fig:architecture} illustrates, ZCs are \emph{topological} aggregators, not domain controllers: a single ZC typically hosts ECUs belonging to different VLANs, and the VLAN boundary is, what limits reachability.
This is least privilege at the link layer.

Eventually, T6 requiring a dynamic and strict access control, is insofar satisfied by the same combination of static VLANs, ACLs, and MACsec that a strict access control is implemented.
The dynamic part is missing, due to the non-existence of reconfiguration during runtime.

Two observations are worth pinning down.
First, the security posture above is achieved without any management plane at runtime.
That is precisely what makes Step~1 deployable right now.
Second, the qualifier \enquote{no SDN at runtime} is not the same as \enquote{no SDN ever}: Step~1 still requires an out-of-band, design-time provisioning step that writes CAKs and VLAN tables into each ECU and switch.
The next subsection makes explicit why the absence of a runtime counterpart to that provisioning step is what bounds Step~1's fulfillment of NIST's ZTA tenets.

\subsection{Remaining Gaps}\label{sec:step1-gaps}

Four gaps remain, that Step~1 cannot fully satisfy:
\begin{compactitem}
  \item[\textbf{G1}] No dynamic policy: VLANs, ACLs, and CAKs are fixed at design time and there is no in-band path to a policy update.
  \item[\textbf{G2}] No standardized and centralised key lifecycle: A CAK lives as long as the ECU it was provisioned into; rotation needs a reflash or workshop visit, and revocation is not a first-class operation. Currently no standardized solution to this exist.
  \item[\textbf{G3}] No runtime visibility or audit: MACsec counters are local; no vehicle-wide instance can query which ECUs are currently authenticated on which associations, nor reconcile that view against an intended policy.
  \item[\textbf{G4}] No identity beyond key possession: TC17~v1.0 equates \enquote{holds the CAK} with \enquote{is the legitimate party} -- no certificate, no revocation list.
\end{compactitem}

Each of these gaps is, on closer inspection, a missing management plane capability: dynamic policy distribution, centralised key management, runtime state queries, and certificate-based identity.
This bridges to Step~2.
A vehicle that wants to close these gaps need not invent ZTA infrastructure from scratch; it needs a management plane.
And, as we will argue in \cref{sec:step2}, the SDN management plane that the zonal vehicle already requires for TSN, MACsec, and VLAN configuration is exactly that management plane.

%% file: step2.tex
\section{Step~2: SDN-Managed Zero Trust}\label{sec:step2}

Step~2 builds on top of Step~1: the MACsec/MKA/VLAN data and control plane is kept verbatim and a CORECONF/YANG-based SDN management plane -- of the kind already on Open Alliance TC19's roadmap for network configuration, e.g. for TSN and QoS configuration -- is added on top.
The key claim of this section is that this management plane, once present, \emph{is} the ZTA control plane: it closes all five Step~1 gaps without replacing any link-layer mechanism and without introducing a second infrastructure dedicated to Zero Trust.

\subsection{System Model Extension}\label{sec:step2-system}

\begin{figure}[t]
\centering
\begin{tikzpicture}[
    font=\sffamily\scriptsize,
    >={Stealth[length=1.6mm]},
    hpc/.style   ={draw, rounded corners=2pt, fill=blue!10, minimum width=20mm, minimum height=7mm, align=center},
    zc/.style    ={draw, rounded corners=2pt, fill=orange!15, minimum width=16mm, minimum height=6mm, align=center},
    ecu/.style   ={draw, rounded corners=1pt, fill=gray!10, minimum width=10mm, minimum height=4.5mm, font=\sffamily\tiny, align=center},
    macsec/.style={line width=0.55pt},
    coreconf/.style={dashed, ->, draw=blue!60!black, line width=0.55pt},
    lane/.style  ={draw, rounded corners=2pt, minimum width=24mm, minimum height=7mm, align=center, font=\sffamily\tiny},
    lanelbl/.style={font=\sffamily\tiny\bfseries, anchor=south west, inner sep=1pt},
    vlanA/.style ={draw=violet, line width=0.7pt},
    vlanB/.style ={draw=teal,   line width=0.7pt},
    vlanC/.style ={draw=brown,  line width=0.7pt},
]
\begin{scope}[local bounding box=topo]
  \node[hpc] (hpc) at (0,0) {VC with SDN Ctrl};
  \node[zc, below left=10mm and 0mm of hpc] (zcf) {ZC-Front\\\tiny SDN Agent};
  \node[zc, below right=10mm and 0mm of hpc] (zcb) {ZC-Body\\\tiny SDN Agent};
  \draw[macsec] (hpc) -- (zcf);
  \draw[macsec] (hpc) -- (zcb);
  \node[ecu, below=8mm of zcf, xshift=-8mm] (f1) {Cam-F};
  \node[ecu, below=8mm of zcf, xshift= 8mm] (f2) {Mirror-FL};
  \node[ecu, below=8mm of zcb, xshift=-8mm] (b1) {Door-R};
  \node[ecu, below=8mm of zcb, xshift= 8mm] (b2) {Brk-RR};
  \draw[macsec, vlanA] (zcf) -- (f1);
  \draw[macsec, vlanB] (zcf) -- (f2);
  \draw[macsec, vlanB] (zcb) -- (b1);
  \draw[macsec, vlanC] (b2) -- (zcb);
  \draw[coreconf] (hpc.west) to[out=180,in=90] ($(zcf.north west)+(-2mm,1mm)$) -- (zcf.north west);
  \draw[coreconf] (hpc.east) to[out=0,  in=90] ($(zcb.north east)+( 2mm,1mm)$) -- (zcb.north east);
  \draw[coreconf] (hpc.west) to[out=210,in=80] ($(f1.north)+(-1mm,2mm)$) -- (f1.north);
  \draw[coreconf] (hpc.west) to[out=225,in=85] ($(f2.north)+( 1mm,2mm)$) -- (f2.north);
  \draw[coreconf] (hpc.east) to[out=-45,in=95] ($(b1.north)+(-1mm,2mm)$) -- (b1.north);
  \draw[coreconf] (hpc.east) to[out=-30,in=100] ($(b2.north)+( 1mm,2mm)$) -- (b2.north);
\end{scope}
\begin{scope}[shift={($(topo.east)+(14mm,-1mm)$)}, local bounding box=planes]
  \node[lane, fill=blue!15] (mgmt) at (0, 1.3) {CORECONF/YANG\\over DTLS};
  \node[lane, fill=blue!8]  (ctrl) at (0, 0.0) {MKA per link};
  \node[lane, fill=green!8] (data) at (0,-1.3) {MACsec~|~gPTP};
  \node[lanelbl] at (mgmt.north west) {Mgmt};
  \node[lanelbl] at (ctrl.north west) {Ctrl};
  \node[lanelbl] at (data.north west) {Data};
  \node[font=\sffamily\tiny, anchor=south, align=center] (sdn) at ($(mgmt.north)+(0,2mm)$) {SDN Controller (runtime)};
  \draw[->] (sdn.south) -- (mgmt.north);
\end{scope}
\end{tikzpicture}\caption{Step~2: SDN management plane overlaid on the Step~1 topology of \cref{fig:architecture}.
  Solid edges and VLAN colours (violet=ADAS, teal=Body, brown=Chassis) are unchanged;
  dashed blue arrows are CORECONF/DTLS sessions, all originating at the SDN controller on the VC and terminating at each SDN agent.
  The previously empty \emph{Mgmt} lane on the right is now driven by the runtime SDN controller.}
\label{fig:architecture-step2}
\end{figure}
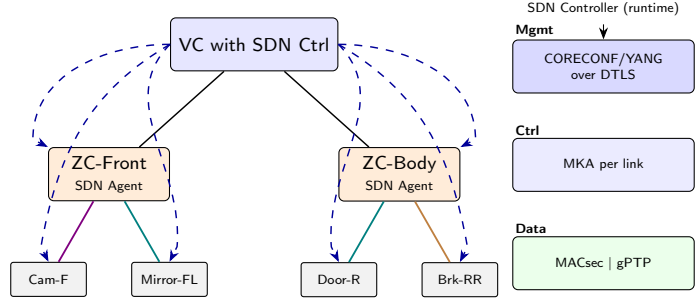

The Step~2 system, see \cref{fig:architecture-step2}, reuses the topology, the connectivity associations, and the static VLAN assignments of Step~1.
Three SDN roles are added on top: the VC additionally hosts an \emph{extended} SDN controller (SDN controller plus the policy logic that turns ZTA decisions into YANG edits), each ZC hosts an SDN agent; the ZC doubles as a Policy Enforcement Point, and each peripheral ECU hosts an SDN agent that receives its own identity material and CAKs and reports local state back to the controller.
With these roles in place the three planes finally separate cleanly: management is CORECONF over DTLS, control remains per-link MKA, and the data plane carries MACsec-protected Ethernet (with e.g. gPTP for time synchronisation).
None of the Step~1 mechanisms are replaced; the empty management lane of \cref{fig:architecture} is simply filled in.

\subsection{Worked Example: Dynamic Reconfiguration}\label{sec:step2-example}

Consider the topology of \cref{fig:architecture-step2} at runtime.
An on-board distributed IDPS hosted on the ZCs and VC observes -- on ZC-Body -- that the rear-right brake actuator \texttt{Brk-RR} is sourcing traffic patterns that no longer match its role; this detection step happens locally on the ZC, outside the SDN planes.
The local IDS sensor forwards the alert to the IDPS instance on the VC, which evaluates the event against the active policy and instructs the SDN controller to quarantine the actuator (\emph{management plane}).
The decision is realised through three small YANG edits pushed over the same CORECONF/DTLS session: revoking the actuator's CAK from the ZC-Body keystore, moving its switch port to a quarantine VLAN, and adding a matching ACL on ZC-Body.
ZC-Body then terminates the affected MKA session (\emph{control plane}), so no further frames from \texttt{Brk-RR} reach the backbone (\emph{data plane}); safety-critical traffic on every other link -- including the entire ADAS path through ZC-Front -- continues unaffected.

\subsection{SDN-to-ZTA Component Mapping}\label{sec:step2-mapping}

The mapping onto NIST SP\,800-207~\cite{nist800207} is direct.
The SDN controller is both Policy Engine -- it evaluates the active policy against the current network state -- and Policy Administrator -- it translates each decision into a YANG configuration edit.
The SDN agent on every ZC and the agent on every peripheral are the Policy Enforcement Points: they own the keys, the MACsec block, and the switch ports through which data-plane traffic must pass.
YANG is the policy language, CORECONF over DTLS is the ZTA management plane, MKA and MACsec remain the enforcement primitives, now under dynamic management.
Trust-algorithm inputs -- IDPS alerts, attestation status, YANG library queries -- reach the controller over the same channel.
NIST's deployment-model taxonomy applies as a hybrid: ZCs are the gateways between zones (\emph{gateway-based}), while the per-link MACsec connectivity associations and VLAN boundaries provide the \emph{micro-segmentation} that a gateway model on its own does not enforce.

A useful sanity check on the \enquote{no new infrastructure} claim is to enumerate the enforcement primitives at the PEP.
There are exactly four, and all four are already required by an SDN-managed automotive Ethernet:
MACsec for per-link authentication and encryption;
VLAN assignment for static and dynamically reconfigurable segmentation, including the quarantine case above;
MKA for session-key derivation, whose teardown is the runtime equivalent of \enquote{revoke access}; and
switch-level ACLs for port- and protocol-granularity filtering.

With this mapping the four Step~1 gaps close as follows:
\begin{inparaenum}
\item[G1] by the runtime CORECONF push of \cref{sec:step2-example};
\item[G2] by the YANG-based key lifecycle of \cref{sec:step2-km};
\item[G3] by YANG library queries that double as the NIST \enquote{asset database};
\item[G4] by binding each agent to a certificate verified during DTLS mutual authentication, with CRL checking centralised on the controller.
\end{inparaenum}
\subsection{YANG-based Network Access Control}\label{sec:step2-nac}

The component mapping above leaves two questions open:
how does an agent first prove it is entitled to talk to the SDN controller at all, and
how does it obtain its first CAK?
The default networking answer is the 802.1X/RADIUS port-based network access control~(PNAC) stack: an EAPoL exchange between the edge node (supplicant) and the switch (authenticator) tunnels EAP through to a back-end RADIUS server, which validates the supplicant's credentials and ships the resulting CAK back to the switch over RADIUS attributes.
We argue that this stack is redundant once the SDN management plane is in place, and can be replaced by a CORECONF/YANG flow that, as we argue below, preserves the authentication and authorisation guarantees it provides.

Three observations motivate the replacement.
\begin{inparaenum}
\item deploying 802.1X / RADIUS \emph{alongside} the SDN management stack duplicates the asymmetric crypto, the certificate handling, and the per-node secure channel that CORECONF / DTLS already requires -- the two stacks solve the same problem twice on the same wire.
\item 802.1X requires each switch to fetch and check CRLs independently, which presupposes connectivity to the central VC, to an off-board back-end or the RADIUS server.
\item today's CAK provisioning lives in OEM-proprietary toolchains; re-expressing it as YANG configuration edits standardises it on infrastructure that becomes part of the vehicle with the SDN management.
\end{inparaenum}

Operationally, the centralised flow has four steps.
The agent authenticates to the controller via DTLS mutual authentication using its X.509 certificate;
the controller checks the certificate and the active policy and decides whether the node is authorised;
on success, the controller distributes the relevant CAK as a YANG configuration edit (wrapped as described in \Cref{sec:step2-km});
standard MKA then derives the SAK and MACsec activates.
Until MACsec is up the switch enforces deny-by-default with one bootstrap exception: CORECONF management frames and MKA control frames are permitted, every other Ethertype is dropped.
This keeps the management and control planes reachable during cold start while the data plane stays silent.
Authorisation can even be implicit: if the controller never distributes a matching CAK, no MKA session can form on that link, and no explicit ACL push is required.

For deployments where the controller is temporarily unreachable -- early startup of a deeply embedded sub-domain, or a multi-hop bootstrap path -- a decentralised fallback based on EAP-TLS~\cite{rfc9190} between edge node and switch can be realised, with the controller pre-distributing credentials and the local authorisation policy in advance.
From MKA's perspective the distributed CAK remains a pre-shared key, so the flow stays within the Open~Alliance TC17~v1.0 PSK recommendation while adding what static \emph{proprietary} PSK provisioning lacks: revocation, runtime rotation, and an auditable key lifecycle.

\subsection{YANG-based Key Management}\label{sec:step2-km}

The NAC flow above assumes that, once the controller authorises a node, a CAK can simply be \enquote{distributed as a YANG configuration edit}.
This subsection makes that step concrete.
To avoid exposing cleartext cryptographic material during transmission or within the volatile memory (RAM) of the application software, we implement a two-stage key lifecycle strategy utilizing \textit{hidden} and \textit{wrapped} key configurations.

\paragraph{Phase 1: Factory provisioning}
During the ECU production phase, a long-term master Key Encryption Key ($\text{KEK}$) is securely programmed directly into a dedicated, write-protected hardware slot of the HSM via a trusted physical interface (e.g., secure JTAG or UDS bootloader routines). 

Crucially, within the ECU's active YANG startup configuration, this key is instantiated as a \textit{hidden key} using the framework specified by the \texttt{ietf-keystore} model~\cite{rfc9642}.
The management plane acknowledges the existence, identifier, and cryptographic algorithm of $\text{KEK}$, but any attempt to read its value via the management protocol yields an empty or null reference. The actual key material never leaves the HSM boundary.

\paragraph{Phase 2: Runtime provisioning}
At runtime, when a Zone Controller establishes a secure MACsec link with another network node via MKA, it requires a shared CAK.

The provisioning protocol flows as $\text{Payload} = \text{E}_{\text{KEK}}(\text{CAK})$, executed in four steps:
\begin{inparaenum}[(1)]
\item the SDN controller generates a random 128- or 256-bit CAK for the MKA session;
\item it encrypts the CAK under the designated $\text{KEK}$;
\item the resulting ciphertext ($\text{encrypted-value}$) plus a reference to the wrapping key (asymmetric-/symmetric-key-ref) is encapsulated into a YANG configuration instance against \texttt{ietf-keystore}~\cite{rfc9642} with cryptographic types drawn from \texttt{ietf-crypto-types}~\cite{rfc9640};
\item the instance is serialised into CBOR and transmitted to the ZC over a secure CORECONF/DTLS session.
\end{inparaenum}

\begin{lstlisting}[language=json, caption={YANG initial key injection (conceptual JSON representation)}, label={lst:yang-init}]
{
  "ietf-keystore:keystore": {
    "symmetric-keys": {
      "symmetric-key": [
        {
          "name": "master-kek-01",
          "hidden": [null]
} ] } } }
\end{lstlisting}

\begin{lstlisting}[language=json, caption={YANG runtime key configuration (conceptual JSON representation)}, label={lst:yang-runtime}]
{
  "ietf-keystore:keystore": {
    "symmetric-keys": {
      "symmetric-key": [
        {
          "name": "mka-cak-zone-1",
          "encrypted-key": {
            "encrypted-by": {
              "symmetric-key-ref": "master-kek-01"
            },
            "encrypted-value-format": "ietf-crypto-types:cms-encrypted-data-format",
            "encrypted-value": "BASE64-CMS-ENCRYPTED-DATA-DER"
} } ] } } }
\end{lstlisting}

\Cref{lst:yang-init,lst:yang-runtime} show the conceptual example representation of the start configuration and runtime payload required to provision the MKA CAK.\footnote{Analogously an asymmetric KEK can be used, following \cite{rfc9640}.}

Upon receiving the CBOR payload, the embedded CORECONF agent parses the data structure and invokes, e.g., the AUTOSAR CSM to execute the key-unwrap routine.
The decryption job is handed over to the HSM, which uses the $\text{KEK}$ stored in its secure internal memory to decrypt the CAK and persists it in its secure key storage.

\emph{Bootstrapping for latency}.
Note that a fresh CORECONF/YANG configuration is not required to be pushed on every power cycle.
Each SDN agent persists its YANG datastore to non-volatile memory and re-reads it locally at boot.
The HSM also persists its secure key storage, so the $\text{KEK}$ and any previously unwrapped CAKs remain available across power cycles.
A cold start therefore reduces to loading the cached configuration, booting the HSM, and running MKA.
Relative to Step~1 this adds no additional operation.
The full DTLS handshake and CORECONF push only occur when something actually has to change: initial commissioning, scheduled key rotation, certificate renewal, or a runtime policy edit triggered by, e.g., \cref{sec:step2-example}.

\subsection{Tenet coverage}
With these flows in pace, NISTs tenets are satisfied as \cref{tab:tenets} summarises.

\begin{table}
\centering
\caption{Step~2 coverage of NIST SP\,800-207 tenets.}
\label{tab:tenets}
\begin{tabular}[t]{lccp{8cm}}
  \toprule
  Tenet & Step~1 & Step~2 & Comment \\
  \midrule
  T1 &       --     &    \CIRCLE   & Addressable YANG resources \\
  T2 &    \CIRCLE   &    \CIRCLE   & MACsec and MKA per link \\
  T3 & \RIGHTCIRCLE &    \CIRCLE   & least-privilege per-session access \\
  T4 &       --     & \RIGHTCIRCLE & trust algorithm for decision logic out of scope \\
  T5 &       --     & \RIGHTCIRCLE & continuous monitoring; no device attestation \\
  T6 & \RIGHTCIRCLE &    \CIRCLE   & dynamic trust algorithm; strictly enforced by MACsec, dynamically per YANG edit \\
  T7 &       --     &    \CIRCLE   & Mgmt channel closes telemetry loop \\
  \bottomrule
\end{tabular}
\end{table}

%% file: conclusion.tex
\section{Conclusion}\label{sec:conclusion}

Does automotive Zero Trust require dedicated infrastructure or can a SDN management plane carry the load?
Our two-step construction gave a lightweight answer.

Step~1 shows that Open~Alliance TC17~v1.0, static MACsec and MKA with pre-shared CAKs extended with VLAN segmentation, already satisfies one NIST SP~800-207 tenet outright (T2, secure communication) and partially two more (T3, least-privilege per-session access; T6, strictly enforcing access), and is deployable today.

Step~2 fills the residual gaps by overlaying a CORECONF/YANG management plane in line with the TC19 direction, in which the SDN controller doubles as Policy Engine and Policy Administrator and every SDN agent on a ZC or peripheral ECU acts as Policy Enforcement Point.
A YANG-based network access-control flow subsumes the 802.1X/RADIUS stack while preserving the authentication guarantees it provides, and a YANG-based key-management scheme distributes CAKs as wrapped values whose decryption never leaves the HSM.

Together the two steps cover five of the seven NIST tenets fully at the network layer (T1--T3, T6, T7) and the remaining two partially (T4, T5) without introducing a single new protocol stack: MACsec, MKA, VLANs, ACLs, DTLS, PKI certificates, CORECONF, and the IETF YANG keystore and crypto-types models are all repurposed rather than reinvented.

Adding a trust algorithm and device attestation to the Step~2 controller will close the remaining gaps. 
These will still make use of the same SDN management plane and thus not introduce any dedicated ZTA infrastructure.

\paragraph{Limitations.}
However, we do not achieve a complete automotive Zero Trust architecture, and the following limitations remain:
\begin{compactitem}
  \item[\textbf{(L1)}] \textbf{Network layer only}:
    We secure links, ports, and connectivity associations; application-layer authorisation inside the ECU is not addressed.
    An ECU authorised on its link may still misbehave within the traffic profile its VLAN and ACLs permit.
  \item[\textbf{(L2)}] \textbf{Identity is delegated to the PKI}.
    Node identity is exactly as strong as the per-ECU certificate assumed in \Cref{sec:bg-threat}, and we do not contribute a provisioning scheme for it (we assume it is injected or generated during production).
  \item[\textbf{(L3)}] \textbf{Per-session authorisation (T3) stays partial}.
    MACsec frames are authorised by possession of a symmetric key, and runtime rotation narrows but cannot close that gap without restrictions on the start up time.
  \item[\textbf{(L4)}] \textbf{No trust algorithm}.
    Step~2 provides inputs, but the interplay and scoring rules that NIST SP~800-207~\S3.3 expects remain to be designed.
  \item[\textbf{(L5)}] \textbf{Resilience is not analysed}.
    Persisted datastores let a vehicle boot and operate with an unreachable controller, so management-plane availability is not on the critical path; controller redundancy, an explicit fail-safe policy, and the degradation logic that must gate revoking a safety-relevant actuator remain open, as it belongs to SDN resilience concepts.
\end{compactitem}
Beyond these, the trust assumptions of \Cref{sec:bg-threat} apply.

Three threads stand out for future work:
\begin{inparaenum}
    \item An end-to-end proof-of-concept on representative automotive hardware; 
    \item A concrete trust algorithm that consumes the management-plane telemetry already collected in Step~2 and emits CORECONF policy edits as its actions; and
    \item An extension of the same YANG models beyond Ethernet, where CANsec on CAN~XL is the natural next link layer to fold in, leaving the authentication, authorisation, and key-provisioning flow unchanged.
\end{inparaenum}

\paragraph{Take-away.}
OEMs need not choose between deployability and Zero Trust.
Step~1 is available now; Step~2 is incremental on top of an SDN management plane that the move to software-defined vehicles requires anyway.
Security, in this construction, comes \emph{by design} rather than \emph{by addition}.